\documentclass[journal]{IEEEtran}
\usepackage{amsmath,amssymb}
\usepackage{graphicx}
\usepackage{booktabs}
\usepackage{xcolor}
\usepackage[hidelinks]{hyperref}

\newcommand{\numModels}{165}

\begin{document}

\title{One Token Is Enough: Fingerprinting and Verifying\\Large Language Models from Single-Token Output Distributions}

\author{Tom\'a\v{s}~Bruckner%
\thanks{T. Bruckner is with the Faculty of Informatics and Statistics, Prague University of Economics and Business, Prague, Czech Republic (e-mail: bruckner@vse.cz).}%
}


\maketitle
\bstctlcite{BSTcontrol}

\begin{abstract}
Large language models (LLMs) are increasingly consumed through opaque
serving chains -- API aggregators, resellers, and inference
providers -- in which the client has no technical means to confirm
that the model answering is the model advertised, and recent audits
show that a substantial fraction of commercial endpoints deviate from
the vendor's reference weights. Existing identification techniques
require long generated texts, token-level log-probabilities,
adversarially crafted prompts, or the model owner's cooperation. We
show that far weaker evidence suffices. We define a behavioral
fingerprint of an LLM as the empirical distribution of its answers to
trivial one-word prompts -- ``name a random number between 1 and
100'' -- collected across four languages at a cost of one output token
per query. Measuring \numModels\ models served via a large commercial
aggregator (OpenRouter), we find that (i)~these distributions are
highly non-uniform (median cell entropy 1.0~bit) and model-specific:
split halves of the same model's samples lie an order of magnitude
closer than samples of different models; (ii)~Jensen--Shannon
divergence between fingerprints recovers model lineage, assigning a
model to its documented family with 59.5\% leave-one-out accuracy
against an 18.4\% chance rate; and (iii)~a biometric-style
verification protocol achieves a 7.3\% equal error rate with the full
40-cell battery, and below 11\% with eight probe cells -- roughly a
hundred single-token queries per audit. We further report ecosystem
anomalies, including a proprietary-branded flagship endpoint
distributionally indistinguishable from an open-weight Qwen model.
The protocol, prompts, raw data, and analysis code are released for
reproduction and operational use.
\end{abstract}


\IEEEpeerreviewmaketitle

\section{Introduction}
\IEEEPARstart{T}{he} market for large language model inference has
rapidly stratified. Between the model creator and the end application
now sit inference providers, resellers, and aggregators that route
requests among dozens of upstream deployments. The client addresses a
model by name -- a string -- and receives text. Nothing in this exchange
proves that the advertised model produced the answer: the provider may
substitute a cheaper model, an aggressively quantized variant, or an
older version, and pocket the margin. This is not a hypothetical
concern. Gao \emph{et al.}~\cite{gao2025model} found that 11 of 31
commercial endpoints serving Llama models produced output distributions
statistically incompatible with the vendor's reference weights; Cai
\emph{et al.}~\cite{cai2025getting} formalize this \emph{model
substitution} threat and show that na\"ive output-based checks are
brittle under production nondeterminism; Zhu \emph{et
al.}~\cite{zhu2025auditing} document the same concern for quantized
variants. Aggregators themselves acknowledge heterogeneous quantization
among their upstream providers.\footnote{OpenRouter, ``How OpenRouter
model routing works,''
\url{https://openrouter.ai/blog/insights/model-routing/}, accessed
2026-07-04.} The economic incentive is structural: inference cost falls
steeply with quantization and model size, while detection risk has so
far been low.

Verifying \emph{which} model is behind an API is therefore a forensic
attribution problem, and it is harder than it looks. The client
typically has (i)~no access to weights or logits -- many production APIs
return text only; (ii)~no ability to fine-tune or watermark the model,
ruling out cooperative techniques~\cite{kirchenbauer2023watermark,
xu2024instructional, shao2026fitprint}; and (iii)~a budget: continuous
auditing of many endpoints must cost cents, not dollars. Existing
non-cooperative identification methods fall short of at least one of
these constraints. Classifier-based attribution of generated
text~\cite{uchendu2020authorship, sun2025idiosyncrasies} needs long
outputs. LLMmap~\cite{pasquini2025llmmap} and
TRAP~\cite{gubri2024trap} achieve excellent accuracy with few queries,
but rely on carefully engineered or adversarially optimized prompts
whose very distinctiveness makes them fragile against an adaptive
provider that recognizes and special-cases them. Statistical equality
testing~\cite{gao2025model} is the closest to our setting but consumes
complete sampled strings and assumes a reference deployment of the
\emph{same} model for calibration.

This paper starts from a different observation, imported from the
behavioral literature on LLMs: when asked for a ``random'' number,
models are reliably, idiosyncratically non-random~\cite{renda2023can,
vankoevering2024random, harrison2024comparison,
coronadoblazquez2025deterministic}. We treat this well-documented defect
not as a curiosity but as a forensic signal. A prompt such as
\emph{``Name a random number between 1 and 100''} costs a few dozen
input tokens and exactly one answer token, yet the \emph{distribution}
of answers over repeated samples is a rich, stable object: it reflects
tokenizer structure, training-data priors, and post-training choices,
none of which a substituted model reproduces exactly. Because the
signal is a distribution over an innocuous everyday question -- not a
magic string -- it is cheap to collect, robust to prompt paraphrasing,
and hard for a dishonest provider to special-case without emulating the
claimed model wholesale.

We develop this observation into a measurement study and a practical
verification protocol. Concretely, we ask:

\begin{itemize}
\item \textbf{RQ1 (Fingerprint existence).} Do distributions of
single-token answers to trivial prompts constitute a stable model
fingerprint -- invariant across serving providers, sampling temperature,
and time, yet distinctive across models?
\item \textbf{RQ2 (Lineage recovery).} Does distance between
fingerprints recover model genealogy, i.e., do models cluster into
their known families, and can an unseen model's family be classified
from its fingerprint alone?
\item \textbf{RQ3 (Verification).} How reliably can a claimed identity
(``this endpoint serves model $X$'') be verified from $k$ probe
queries, measured as a biometric-style ROC/EER, and how does
reliability scale with the query budget?
\item \textbf{RQ4 (Ecosystem audit).} Applied to a production
aggregator ecosystem, what anomalies does the fingerprint reveal -- 
endpoints whose behavior diverges from their declared model?
\end{itemize}

\subsection{Contributions}
\begin{enumerate}
\item \textbf{A minimal-output fingerprint.} We define and validate a
black-box behavioral fingerprint of an LLM requiring only one output
token per query and no logits, no long generations, and no cooperation
from the model owner (Sec.~\ref{sec:method}).
\item \textbf{The largest single-token behavioral census of served
models to date.} \numModels\ models probed through a commercial
aggregator with a battery of 10 fingerprinting tasks in four languages
(English, Russian, Chinese, Arabic), $\sim$326{,}000~requests,
total data-collection cost under \$35
(Sec.~\ref{sec:setup}).
\item \textbf{A verification protocol with an explicit cost--reliability
curve.} Split-sample Jensen--Shannon divergence yields an equal error
rate of 7.3\% with the full battery and 10.6\% with eight probe cells
($\sim$120 single-token queries), quantifying, for the first time for
this class of methods, how many tokens a trustworthy identity check
costs (Sec.~\ref{sec:results}).
\item \textbf{Ecosystem findings and artifacts.} Anomalies surfaced
by the fingerprint -- a proprietary-branded flagship endpoint
distributionally indistinguishable from an open-weight Qwen model,
and 10 of 34 same-model provider pairs diverging beyond the impostor
range -- presented as independently verifiable distributional
deviations (Sec.~\ref{sec:ethics}); all prompts, raw responses with
full serving metadata, and analysis code are released.
\end{enumerate}

The fingerprint's multilingual design additionally captures
\emph{shared cultural priors} across models -- a property we exploit in
a companion study of tacit coordination between models; here we use
languages only as independent probe dimensions and defer the
coordination analysis to future work.

\section{Related Work}\label{sec:related}

\subsection{Model Fingerprinting and Attribution}
Attribution of generative-model outputs is an established forensic
problem in TIFS and adjacent venues, spanning synthetic-image
attribution~\cite{wang2025bosc, zheng2026adaparse} and forged-speech
attribution~\cite{zhang2025softcontrastive}; our work is the
text-domain analogue under the tightest possible output budget. For
LLMs specifically, \emph{cooperative} (white-box) schemes embed
verifiable marks: decoding-time
watermarks~\cite{kirchenbauer2023watermark}, instruction-tuned
backdoors~\cite{xu2024instructional}, and false-claim-resistant
ownership fingerprints~\cite{shao2026fitprint}. These serve the model
\emph{owner}; a third-party client auditing an untrusted endpoint can
deploy none of them. \emph{Non-cooperative} identification includes
authorship attribution of machine text~\cite{uchendu2020authorship},
which extends to modern LLMs: Sun \emph{et
al.}~\cite{sun2025idiosyncrasies} attribute full chat responses to
their source model with 97\% five-way accuracy, showing that strong
identity signal exists in text -- but their classifiers consume long
generations. Closest in spirit are query-efficient active probes:
LLMmap~\cite{pasquini2025llmmap} identifies 42 model versions with
eight engineered queries, and TRAP~\cite{gubri2024trap} repurposes
adversarial suffixes as identity honeypots. Both optimize \emph{which
strings} to send; we instead optimize \emph{what to measure}, reading
identity off the response distribution to fixed, innocuous,
paraphrasable prompts. This trades single-shot identification for
robustness: our probes are indistinguishable from ordinary traffic,
and the signal survives provider-side prompt inspection
(Sec.~\ref{sec:adaptive}). Backdoor-detection work within this venue
likewise treats model behavior under crafted inputs as forensic
evidence~\cite{wei2024bdmmt}.

\subsection{API Auditing and Model Substitution}
Gao \emph{et al.}~\cite{gao2025model} formalize \emph{model equality
testing} as a two-sample test on sampled strings and provide the
central empirical motivation for this line of work: a third of audited
Llama endpoints deviated from reference weights. Cai \emph{et
al.}~\cite{cai2025getting} taxonomize substitution attacks
(quantization, smaller models, version rollback) and show that simple
output comparisons fail under benign nondeterminism, motivating
hardware attestation; Zhu \emph{et al.}~\cite{zhu2025auditing} detect
quantized substitutes with a rank-based uniformity test assuming
log-probability access. Our protocol complements this literature on
two axes: it needs neither full strings nor log-probabilities (text-only
endpoints suffice), and it is calibrated \emph{across} models rather
than against a single reference deployment, which lets the same data
answer attribution (which family?) and verification (this model?)
questions.

\subsection{Behavioral Regularities of LLMs}
That LLMs cannot sample uniformly is well documented: biased
``random'' numbers~\cite{renda2023can, harrison2024comparison},
human-like and amplified sequence biases in coin
flips~\cite{vankoevering2024random}, and model- and language-dependent
favorite numbers~\cite{coronadoblazquez2025deterministic}. This
literature treats non-randomness as a deficiency to characterize or
correct. Behavioral-economics studies of LLMs similarly document
systematic, model-specific dispositions in games~\cite{akata2025playing},
and a parallel line shows that shared training corpora induce shared
priors and homogenized outcomes across
deployments~\cite{bommasani2022picking, kleinberg2021algorithmic},
echoing classic focal-point behavior in
humans~\cite{schelling1960strategy, mehta1994nature}. We contribute the
inverse perspective: whatever is systematic and model-specific in these
biases is \emph{identity}, and identity that costs one token to sample
is forensically valuable. To our knowledge, no prior work uses
single-token answer distributions for model identification or
verification at ecosystem scale.

\section{Threat Model}\label{sec:threat}
\textbf{Setting.} A \emph{client} pays for inference from a declared
model $X$ through an untrusted \emph{provider} (possibly reached via an
aggregator that routes among providers). The provider controls the
serving stack end-to-end: it may serve $X$ faithfully, serve a
quantized or otherwise degraded variant $X'$, substitute a cheaper
model $Y$, or route dynamically among these.

\textbf{Verifier capabilities.} The client (or an auditor acting for
many clients) can (i)~send ordinary chat requests with a chosen system
and user prompt, temperature, token limit, and standard decoding
options (e.g., disabling optional reasoning modes); (ii)~observe returned
text and standard metadata; and (iii)~query a \emph{reference}
deployment of $X$ that it trusts (e.g., the vendor's first-party API)
to enroll a reference fingerprint. The verifier has no logits, no
weights, and a per-audit budget of at most a few hundred output tokens.

\textbf{Adversary capabilities and goals.} The economically rational
adversary minimizes serving cost while avoiding detection. We consider
three tiers: (T1)~\emph{oblivious} -- substitutes silently, does not
inspect traffic; (T2)~\emph{filtering} -- recognizes known audit prompts
(e.g., published verbatim) and routes them to the genuine $X$;
(T3)~\emph{emulating} -- attempts to reproduce $X$'s answer
distributions on arbitrary low-entropy prompts. Our protocol targets T1
fully and T2 by design (the probe space is an open-ended paraphrase
family of everyday questions, sampled at audit time;
Sec.~\ref{sec:adaptive}); T3 collapses into faithfulness, since
matching $X$'s conditional output distributions on arbitrary prompts is
tantamount to running $X$.

\section{Method}\label{sec:method}

\subsection{Fingerprint Definition}
Let $M$ be a model served at some endpoint, and let
$\mathcal{B}=\{(t,\ell)\}$ be a battery of \emph{probe cells}: task $t$
(e.g., \emph{random number 1--100}) crossed with prompt language
$\ell$. For each cell, the endpoint is queried $n$ times at temperature
$1.0$ with a fixed system prompt enforcing a one-word answer, a hard
cap of 16 completion tokens, and provider-side reasoning modes
disabled, so that the completion is a direct single-pass sample from
the model's conditional distribution.\footnote{The pre-registration
specified a 12-token cap; we raised it to 16 (the minimum accepted by
several vendors' chat endpoints) after a cap of 12 was rejected
outright by those endpoints. The change cannot affect any measured
distribution, as valid answers are one to six tokens; it only lets the
affected models respond. We also excluded, on the pre-registered
mandatory-reasoning grounds, models whose reasoning phase cannot be
disabled at all (the OpenAI o-series reject a reasoning effort of
``none''), which the catalog metadata did not flag. Collection
surfaced two further empirical variants of the same defect, likewise
excluded: endpoints whose serving providers ignore the
reasoning-disable flag and burn the token budget on a reasoning trace,
and flagship chat endpoints whose completions consume $\sim$40--60
tokens per one-word visible answer without exposing any reasoning
trace -- an unverifiable hidden computation phase
(Sec.~\ref{sec:limitations}). The token cap is 16 for the entire
census; per-model exclusion reasons ship with the artifact.} Answers are normalized (case folding,
punctuation stripping, digit-system mapping, language-specific
canonicalization; Sec.~\ref{sec:processing}) into a categorical
variable, yielding an empirical distribution
$\hat{p}^{M}_{t,\ell}$. The \emph{fingerprint} of $M$ is the tuple
$F(M)=\big(\hat{p}^{M}_{t,\ell}\big)_{(t,\ell)\in\mathcal{B}}$,
accompanied by a deterministic variant collected at temperature $0$.

\emph{Unit of analysis.} Aggregator catalog identifiers map
many-to-one onto model \emph{checkpoints}: the same weights are
routinely exposed as a rolling alias (\texttt{-latest}), a dated
snapshot, a sponsored free tier (\texttt{:free}), and inference-mode or
serving-tier variants (reasoning effort, fast serving, extended
context, thinking toggles). Throughout, the object being fingerprinted
is the \emph{checkpoint}; an \emph{endpoint} is one serving path to a
claimed checkpoint; a \emph{family} is a lineage of checkpoints sharing
ancestry. Fingerprints are estimated per checkpoint (duplicate
identifiers are removed at selection time, Sec.~\ref{sec:setup});
verification (RQ3) compares an endpoint against a reference deployment
of its claimed checkpoint; lineage analyses (RQ2) operate on families.

The battery (Table~\ref{tab:battery}) contains ten tasks in four
languages (English, Russian, Chinese, Arabic): random and favorite
numbers, random letters, words, colors (random and favorite), animals,
cities, and a coin flip. Tasks were chosen to (i)~admit one-token
answers; (ii)~span closed (numbers, coin) and open (words, cities)
answer spaces, whose distributions carry complementary signal;
(iii)~include the random/favorite contrast, itself
discriminative~\cite{coronadoblazquez2025deterministic}; and
(iv)~be culturally and linguistically neutral enough to pose in all
four languages. Languages multiply the probe dimensions at no design
cost and probe different slices of the training prior.

\begin{table}[t]
\caption{Probe battery (fingerprinting tasks; full prompts in all four
languages in the released artifact, config/prompts.json).}
\label{tab:battery}
\centering
\begin{tabular}{lll}
\toprule
Task & Answer space & Condition \\
\midrule
random number 1--100 & closed (100) & random \\
random number 1--10 & closed (10) & random \\
favorite number & open (numeric) & favorite \\
random letter & closed (alphabet) & random \\
random word & open & random \\
random color & open (canonicalized) & random \\
favorite color & open (canonicalized) & favorite \\
random animal & open & random \\
random city & open & random \\
coin flip & closed (2) & random \\
\bottomrule
\end{tabular}
\end{table}

\subsection{Data Processing}\label{sec:processing}
Raw completions are normalized deterministically: Unicode NFC,
punctuation and quotation stripping, case folding, Arabic-Indic and
Chinese numeral mapping to Latin digits, first-token extraction, and a
per-language color lexicon mapping color terms to canonical codes for
the cross-language analysis. Every response is classified as
\emph{valid}, \emph{invalid} (off answer space or multi-word),
\emph{refusal}, or \emph{empty}; nothing is discarded silently, and
per-model validity rates are reported (overall validity 97.6\%;
per-model median 99.6\%, minimum 60.6\% for a model that frequently
answers in sentences; full table in the released artifact). Refusal
and validity rates
are themselves weak identity signals but are excluded from the
fingerprint to keep it robust to safety-layer changes.

\subsection{Distance and Lineage Analyses}\label{sec:distance}
The distance between two models is the Jensen--Shannon divergence
(base~2) between their distributions, averaged over all cells where
both have at least 10 valid samples:
\begin{equation}
D(M_a,M_b)=\frac{1}{|\mathcal{B}'|}\sum_{(t,\ell)\in\mathcal{B}'}
\mathrm{JSD}\!\left(\hat{p}^{M_a}_{t,\ell}\,\|\,\hat{p}^{M_b}_{t,\ell}\right).
\end{equation}
JSD is symmetric, bounded in $[0,1]$, and well-defined on disjoint
supports -- appropriate for sparse categorical distributions. For RQ2 we
(i)~cluster the distance matrix with average-linkage (UPGMA)
hierarchical clustering and compare the tree against known model
families via the adjusted Rand index at the family-count cut, and
(ii)~classify each model's family by its nearest neighbor,
leave-one-out. Family labels derive from developer documentation of
base-model lineage, curated manually and released with the data;
misclassified models are examined individually as candidate anomalies
(RQ4).

\subsection{Verification Protocol}\label{sec:verification}
Verification asks: \emph{does endpoint $E$ serve claimed model $X$?}
Enrollment: the verifier collects a reference fingerprint of $X$ from a
trusted deployment. Audit: the verifier samples $k$ probe cells from
the battery (or its paraphrase family), collects $n$ answers per cell
from $E$, and computes the score
$s = \frac{1}{k}\sum \mathrm{JSD}(\hat{p}^{\mathrm{ref}}\,\|\,\hat{p}^{E})$,
accepting the claim iff $s\le\tau$. We evaluate this biometric-style:
\emph{genuine} trials compare two disjoint halves of the same model's
samples (split by repetition parity, hence time-interleaved);
\emph{impostor} trials compare model $X$'s half against model $Y$'s
half, $Y\neq X$. Sweeping $\tau$ yields ROC curves, area under the
curve, and the equal error rate (EER); resampling random $k$-subsets of
cells yields the \emph{query-budget curve}, EER as a function of the
number of probe cells -- the operational cost of a given assurance
level. Substituting for genuine trials the same model served by
\emph{different providers} (which OpenRouter's routing supplies as a
natural experiment) tests robustness of the fingerprint to serving
stacks (RQ1); temperature-0 fingerprints are analysed analogously.

\subsection{Implementation}
Data collection runs against the OpenRouter aggregator API with a
resumable, idempotent runner: every request is a deterministic cell
repetition; failures are retried with exponential backoff and never
enter the data; each response is stored verbatim with UTC timestamp,
latency, serving provider, reported model string, token usage
(including cached-token counts, enabling the response-caching screen of
Sec.~\ref{sec:caching}), and cost. Requests disable provider-side
reasoning modes (Sec.~\ref{sec:method}). Cells execute in
seeded-shuffled order, so rate-limit shortfalls
spread approximately uniformly across a model's cells
(missing-at-random by design) instead of depleting whole tasks or
languages; per-cell analyses require at least 10 valid samples
(Sec.~\ref{sec:distance}), and per-model attrition is reported: five
catalog-listed endpoints were unreachable throughout the collection
window and one failed systematically (Sec.~\ref{sec:limitations}); of
the 165 census models, 163 reached $\geq$95\% of expected repetitions
and two (91--94\%) retained all 40 cells above the per-cell minimum.
Prompts, configuration, raw responses with full serving metadata, and
all derived results are released as an open dataset on
Zenodo~\cite{bruckner_2026_21278557}; the complete Node.js/R collection
and analysis pipeline is archived
separately~\cite{bruckner_2026_21278793}.

\section{Experimental Setup}\label{sec:setup}
\textbf{Models.} From the aggregator catalog snapshot of 2026-07-06
(342~models), we exclude, by documented machine-checkable rules,
identifiers that are (i)~not plain text-to-text chat models (non-text
output; embedding/audio/image/video; moderation and guard models; base
non-instruction-tuned models; context below 2{,}048 tokens);
(ii)~narrowly specialized models (code, math, medical, roleplay);
(iii)~not a single stable checkpoint in the sense of
Sec.~\ref{sec:method}: rolling \texttt{-latest} aliases, mutable
preview and deprecated endpoints, meta-routers, multi-agent or
search-grounded pipelines whose output is not a pure model prior, and
free-tier or inference-mode duplicates of checkpoints already in the
catalog; (iv)~endpoints with a \emph{mandatory} hidden reasoning phase,
for which the direct single-pass completion that defines the
fingerprint (Sec.~\ref{sec:method}) is unobservable; (v)~free-tier-only
endpoints, whose account-level daily request caps make full collection
operationally infeasible. One further exclusion is manual and
documented (a model priced at $\$150$/$\$600$ per 1M input/output
tokens). The rule list ships with the artifact and every excluded
identifier carries its machine- or reviewer-assigned exclusion reason.
This leaves \numModels\ models across 19 family labels and
53 serving providers. \textbf{Sampling.} Each model $\times$ 10
tasks $\times$ 4 languages is sampled 30 times at $T{=}1.0$ and
3 times at $T{=}0$; frontier-priced models ($\geq\$5$ per 1M input
tokens) use 15 repetitions at $T{=}1.0$.
\textbf{Pilot and pre-registration.} A pilot over 14 models spanning
seven co-family large+small pairs fixed all design and analysis choices
against four pre-specified criteria: per-model validity $\geq 80\%$,
within-provider greedy determinism $\geq 90\%$ of cells, a significant
inter- vs intra-family divergence gap (permutation test,
$p=0.0008$), and cost extrapolation. Hypotheses and the analysis plan
were pre-registered~\cite{bruckner2026prereg} before the main run,
with pilot data excluded from confirmatory analyses. \textbf{Cost.} The
complete census comprises 326{,}047 responses (23.3M input, 1.16M
output tokens) and cost \$34.44, i.e., \$0.21 per model -- about three
orders of magnitude below long-generation attribution at equal
repetition counts.

\section{Results}\label{sec:results}
Every number below is produced by a named, released artifact file and
is regenerable end-to-end from the raw responses.

\subsection{RQ1: Fingerprints Exist and Are Stable}
Single-token answer distributions are strongly non-uniform: across the
6{,}572 census cells with at least 10 valid samples, the median cell
entropy is 1.00~bit and the median modal-answer share is 0.71, against
uniform-baseline entropies of 1--6.6 bits depending on the task's
answer space (e.g., 6.64 bits for \emph{random number 1--100}). They
are also model-specific and reproducible: the median cell-level JSD
between two disjoint halves of the same model's samples is 0.075,
versus 0.489 between halves of \emph{different} models (6{,}564
genuine vs.\ 1.07M impostor cell-level trials); aggregated over the
battery, this separation yields the verification performance of
Sec.~RQ3. At temperature~0, 90.4\% of cells are deterministic within
serving provider; pooled across providers the rate drops to 84.5\%,
the gap being serving-stack variance -- different upstream
deployments of the same model produce different, individually stable,
greedy answers (cf.\ Sec.~\ref{sec:caching}).

\begin{figure}[t]
\centering
\includegraphics[width=\linewidth]{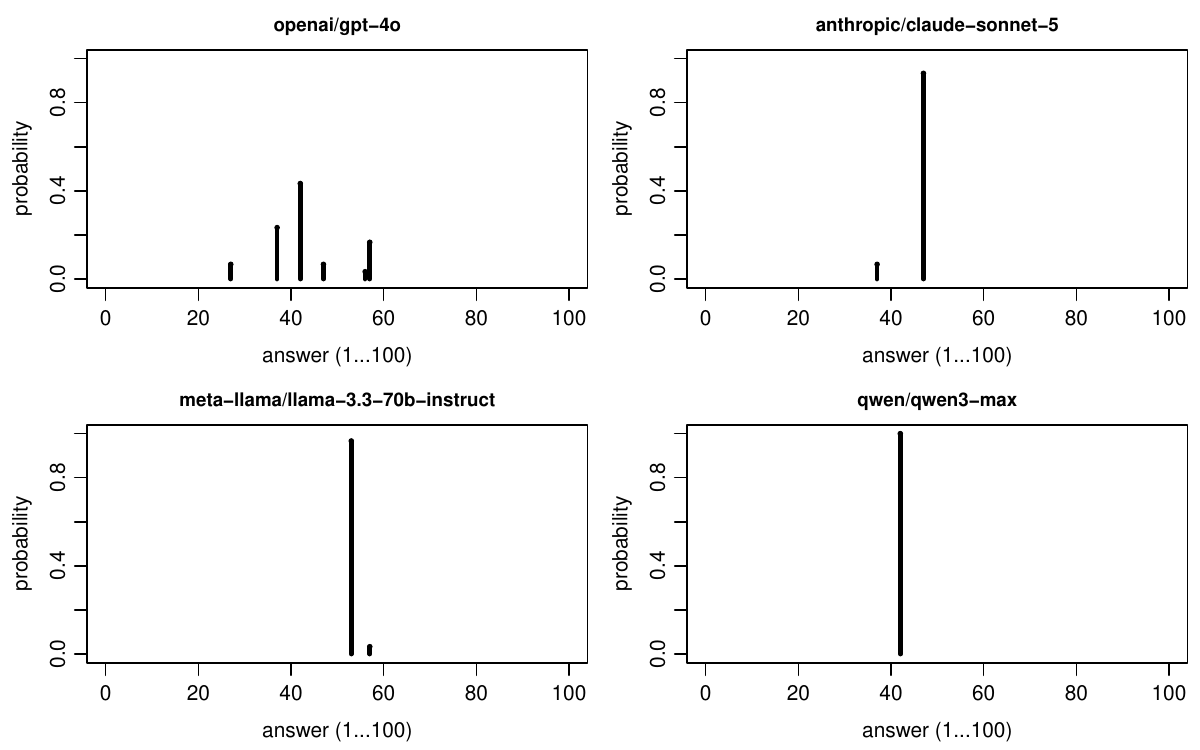}
\caption{The raw fingerprint signal: answer distributions for
\emph{``Name a random number between 1 and 100''} (English, $T{=}1.0$,
30 samples). Same prompt, four models, four sharply different -- and
individually stable -- distributions: GPT-4o spreads over
42/37/57, Claude Sonnet~5 concentrates on 47, Llama~3.3 on 53, and
Qwen3-Max answers 42 every single time.}
\label{fig:examples}
\end{figure}

\emph{Robustness to the serving stack.} OpenRouter's routing provides
a natural experiment: 31 census models were served by two or more
providers with enough coverage to compare deployments directly (34
provider pairs, $\geq$15 shared cells each). The median cross-provider
distance is 0.227 -- above the split-half noise floor of a single
deployment (0.140) but well below the impostor median (0.463), and
70.6\% of provider pairs stay below the 5th percentile of impostor
distances. Verifying a model against a reference collected at a
\emph{different} provider still achieves $\mathrm{AUC}=0.880$ (versus
0.971 same-mix), so the fingerprint largely survives serving-stack
changes; the pairs that do not are precisely the anomaly candidates of
Sec.~RQ4. Fig.~\ref{fig:examples} illustrates the raw signal.

\subsection{RQ2: Lineage Recovery}
\begin{figure*}[!p]
\centering
\includegraphics[width=0.98\textwidth]{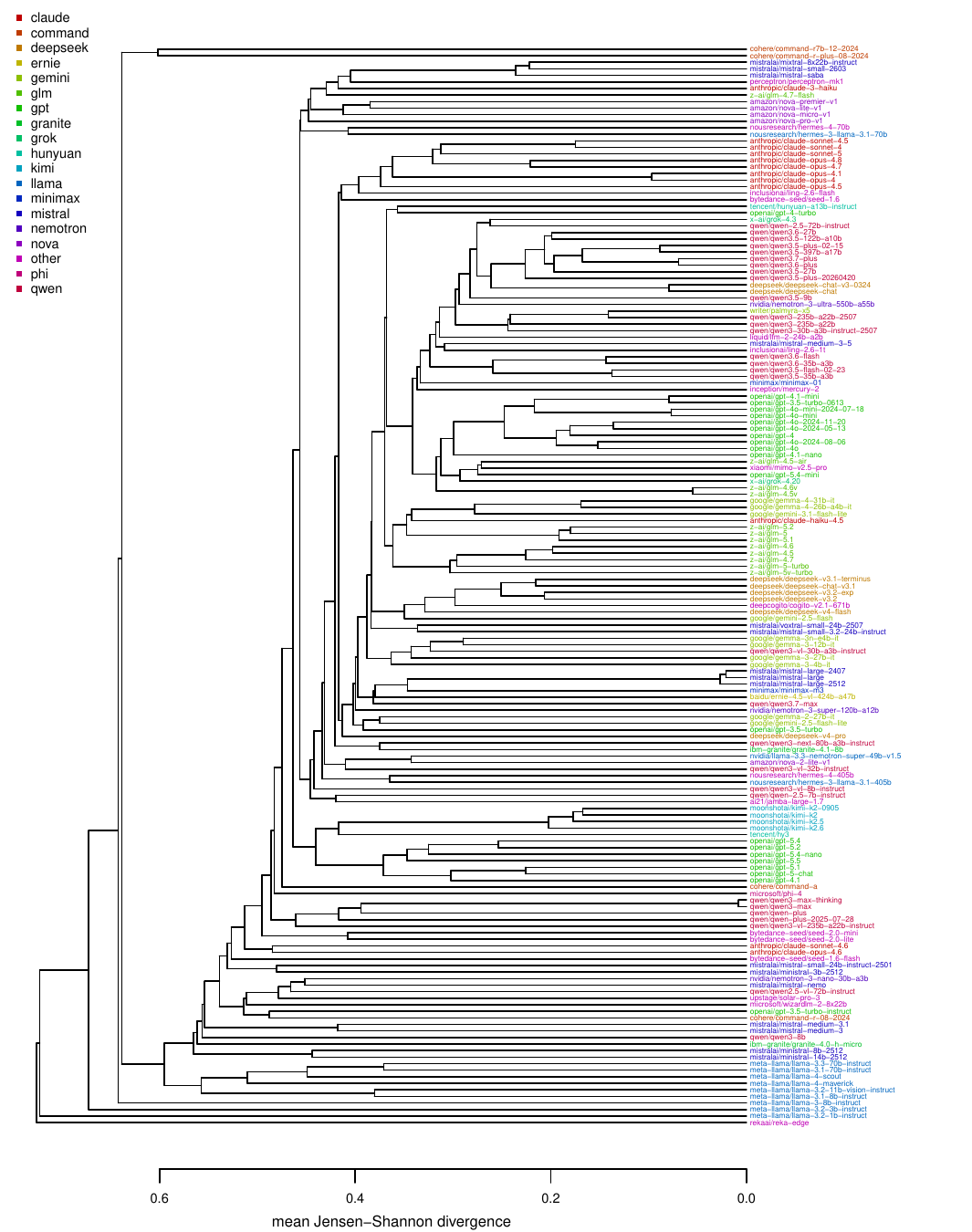}
\caption{Hierarchical clustering (UPGMA) of 165 served models on mean
Jensen--Shannon divergence between single-token fingerprints; leaf
labels are colored by documented family. Labels are small in print but
fully zoomable in the electronic version; a large-format rendering
ships with the released artifact~\cite{bruckner_2026_21278557}.}
\label{fig:dendrogram}
\end{figure*}

Fig.~\ref{fig:dendrogram} shows average-linkage hierarchical
clustering of the census on mean fingerprint JSD; the tree represents
the distance structure faithfully (cophenetic correlation 0.886).
Family structure in fingerprint space is strongly \emph{local}:
leave-one-out 1-NN classification assigns a model to its documented
family with 59.5\% accuracy over the 163 models that have at least one
same-family peer, against an 18.4\% frequency-weighted chance rate
(3.2$\times$ chance; exact binomial $p<10^{-30}$). A flat cut of the
dendrogram at $k=19$ recovers families only weakly (ARI $=0.023$),
which is expected: the label set includes a 17-model ``other''
grab-bag of unrelated vendors and several singletons, and lineage
signal lives in nearest-neighbour geometry rather than in globally
compact clusters. Table~\ref{tab:families} gives per-family
precision/recall; documented major lineages are recovered well
(GLM $1.00/0.83$, Mistral $0.87/0.81$, GPT $0.70/0.90$), while the
heterogeneous ``other'' label predictably is not ($0.20/0.06$).

\begin{table}[t]
\caption{LOO 1-NN family classification, families with $\geq 8$
members (precision/recall).}
\label{tab:families}
\centering
\begin{tabular}{lccc}
\toprule
Family & $n$ & Precision & Recall \\
\midrule
qwen & 30 & 0.50 & 0.73 \\
gpt & 21 & 0.70 & 0.90 \\
other & 17 & 0.20 & 0.06 \\
mistral & 16 & 0.87 & 0.81 \\
claude & 12 & 0.54 & 0.58 \\
glm & 12 & 1.00 & 0.83 \\
llama & 12 & 0.88 & 0.58 \\
gemini & 11 & 0.43 & 0.55 \\
deepseek & 8 & 0.46 & 0.75 \\
\bottomrule
\end{tabular}
\end{table}

\subsection{RQ3: Verification and Query Budget}
\begin{figure}[t]
\centering
\includegraphics[width=0.9\linewidth]{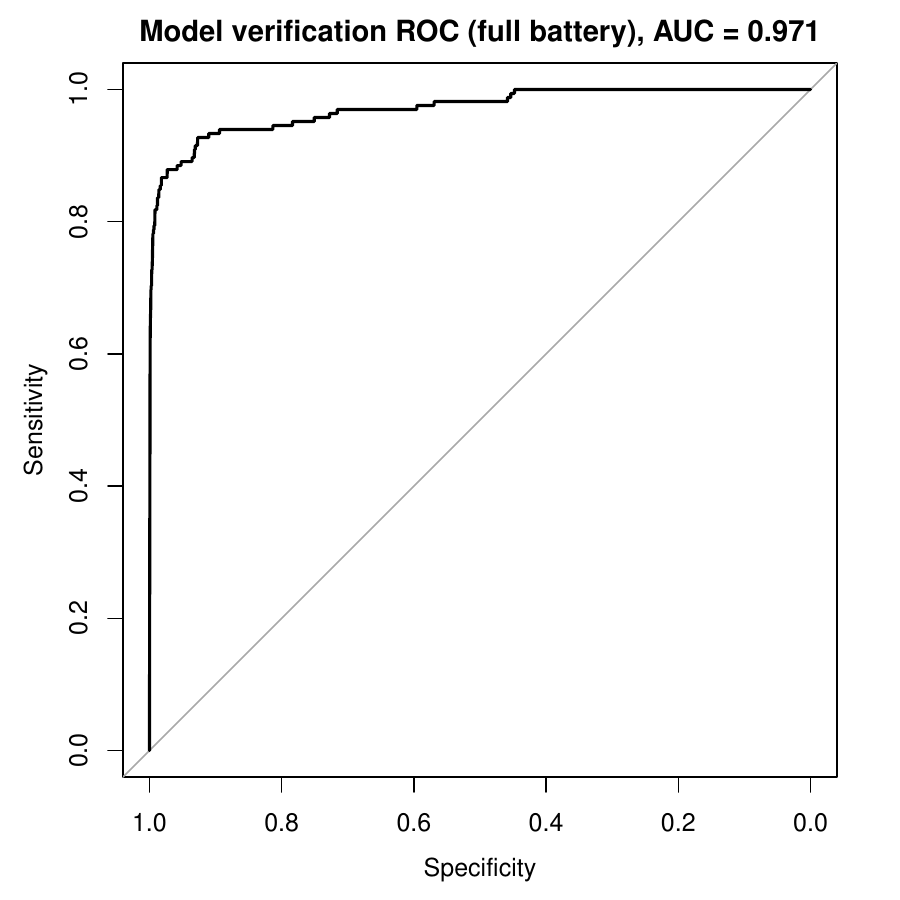}
\caption{Verification ROC: split-half fingerprint distance separates
genuine from impostor identity claims ($\mathrm{AUC}=0.971$,
$\mathrm{EER}=7.3\%$; 165 genuine / 27{,}060 impostor trials).}
\label{fig:roc}
\end{figure}

\begin{figure}[t]
\centering
\includegraphics[width=0.9\linewidth]{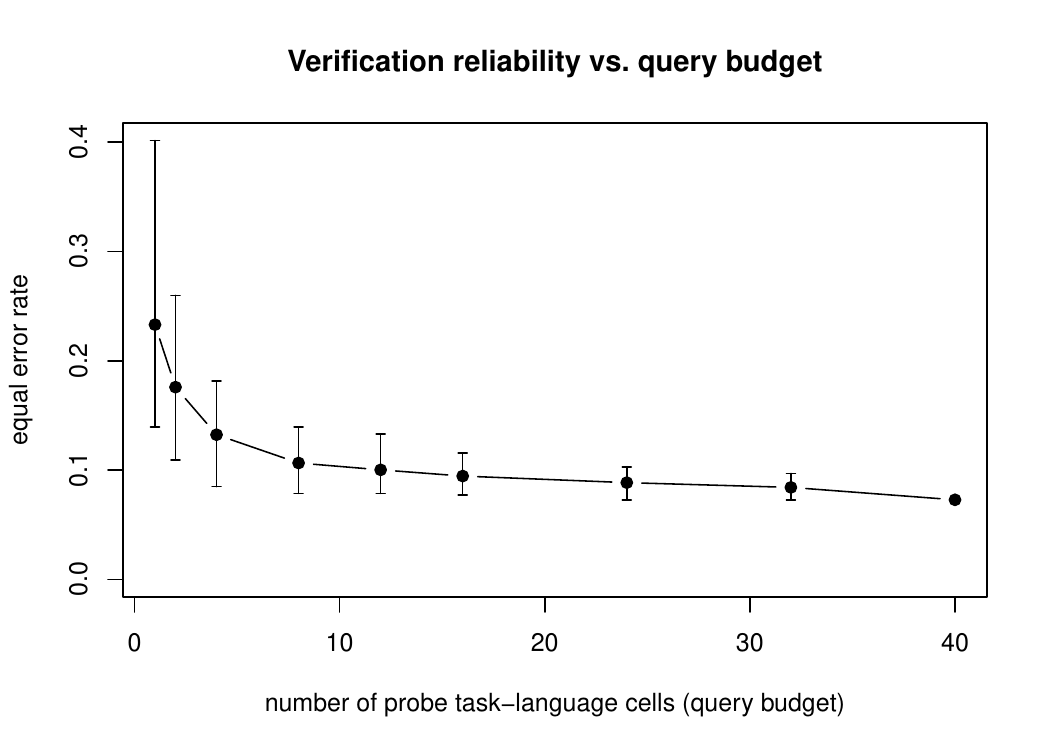}
\caption{Reliability--cost trade-off: EER as a function of the number
of probe cells $k$ (random subsets of the 40-cell battery).}
\label{fig:budget}
\end{figure}

Fig.~\ref{fig:roc} shows the verification ROC over 165 genuine and
27{,}060 impostor full-battery trials: $\mathrm{AUC}=0.971$,
$\mathrm{EER}=7.3\%$. Fig.~\ref{fig:budget} and
Table~\ref{tab:budget} quantify the operational trade-off by
resampling random $k$-subsets of the 40 probe cells: a single cell
already verifies far better than chance ($\mathrm{EER}=23.3\%$), eight
cells reach $10.6\%$, and the curve flattens beyond
$k\!\approx\!16$--24. At the $k{=}16$ operating point, an audit at the
sample size underlying these estimates (15 repetitions per cell, as in
the split halves) costs $16\times15=240$ single-output-token queries,
$\sim$16k input tokens in total: from hundredths of a cent for
typically priced models to a few cents for frontier-priced ones  -- 
continuous re-auditing of an endpoint is economically trivial.

\begin{table}[t]
\caption{Query-budget curve: EER when only $k$ of the 40 probe cells
are used (mean and 90\% band over random subsets).}
\label{tab:budget}
\centering
\begin{tabular}{lcccccc}
\toprule
$k$ & 1 & 4 & 8 & 16 & 32 & 40 \\
\midrule
EER (\%) & 23.3 & 13.2 & 10.6 & 9.5 & 8.4 & 7.3 \\
band (\%) & 14--40 & 9--18 & 8--14 & 8--12 & 7--10 & -- \\
\bottomrule
\end{tabular}
\end{table}

\subsection{RQ4: Ecosystem Anomalies}
All findings below are statistical statements about served
distributions, not claims of intent; benign explanations and
disclosure handling are discussed in Sec.~\ref{sec:ethics}.

\textbf{Identity anomalies.} The strongest case concerns a flagship
endpoint marketed as a vendor's proprietary in-house model
(\texttt{writer/palmyra-x5}): its fingerprint lies at distance 0.141
from \texttt{qwen/qwen3-235b-a22b-2507} -- \emph{at the median of
genuine same-model distances} (0.140) and far below the impostor
range -- i.e., distributionally indistinguishable from an open-weight
Qwen deployment. Conversely, the fingerprint confirms documented but
unlabeled lineage: \texttt{deepcogito/cogito-v2.1-671b}, trained from
a DeepSeek~V3 base, lands nearest to three DeepSeek checkpoints
(0.268--0.308, below the 5th percentile of impostor distances), and
several further ``other''-vendor models
(\texttt{xiaomi/mimo-v2.5-pro}, \texttt{inclusionai/ling-2.6-1t})
place firmly inside the Qwen neighbourhood. The signal can also be
\emph{erased}: \texttt{nvidia/llama-3.3-nemotron-super-49b-v1.5},
a documented Llama-3.3 derivative, is closer to Qwen models (0.303)
than to any Llama checkpoint -- heavy post-training overwrites
lineage priors, which bounds what family classification can promise.

\textbf{Deployment anomalies.} Among the 34 same-model provider pairs
of Sec.~RQ1, ten (29\%) diverge beyond the 5th percentile of impostor
distances -- the served distributions differ more than two distinct
models typically do. The extreme case is
\texttt{meta-llama/llama-3.2-3b-instruct} at Cloudflare vs.\ Parasail
(0.716, deep in impostor territory); notably, even
\texttt{openai/gpt-4} served via Azure vs.\ OpenAI first-party reaches
0.392. Benign explanations for every case, and how these findings are
framed and released, are discussed in Sec.~\ref{sec:ethics}.

\textbf{Serving-layer opacity.} Collection itself surfaced anomalies
invisible to content-blind audits: flagship chat endpoints consuming
$\sim$40--60 completion tokens per one-word visible answer,
providers silently ignoring the reasoning-disable flag (0.76\% of
responses carried a reasoning trace; 14 model--provider combinations),
and one model whose sole provider returned server errors on
\emph{every} numeric prompt while answering all others normally
(Sec.~\ref{sec:limitations}).

\section{Discussion}\label{sec:discussion}

\subsection{Why Single-Token Distributions Carry Identity}
The distribution of a one-token answer to ``name a random color''
aggregates several layers of a model's construction: tokenizer
segmentation of candidate answers, pre-training corpus frequencies,
instruction-tuning and preference-optimization shifts, and decoding
implementation. Quantization and distillation perturb precisely the
final softmax landscape that these probes sample, which is why the
fingerprint is sensitive to the substitutions that matter commercially
while remaining largely insensitive to serving-stack details that do
not: empirically, replacing the serving provider moves a model's
fingerprint by a median of 0.227 -- less than half the impostor
median of 0.463 and enough to keep cross-provider verification at
$\mathrm{AUC}=0.880$ (Sec.~\ref{sec:results}) -- while the minority
of provider pairs that exceed the impostor range are exactly the
deployments our protocol is meant to flag.

\subsection{The Adaptive Provider}\label{sec:adaptive}
A T2 adversary that special-cases published audit prompts faces an
open-ended paraphrase family: by construction, the battery contains no
magic strings -- its tasks are semantic classes (\emph{name a random
color/number/city}) that admit unbounded rewordings and translations,
and the fingerprint is a property of the answer distribution, not of
any particular prompt wording (our own battery already realizes each
task in four languages). A dedicated paraphrase-invariance experiment
is left to future work. Audit prompts are sampled at audit time and
interleavable with ordinary traffic, making filtering
detection-theoretically expensive. A T3 adversary must match
conditional answer distributions on arbitrary everyday prompts;
distillation to that fidelity on the probe family's support
approximates running the claimed model, eroding the substitution
margin. We do not claim cryptographic guarantees: our protocol raises
the cost of undetected substitution, complementing attestation-based
proposals~\cite{cai2025getting}.

\subsection{Serving-Layer Caching}\label{sec:caching}
Repeatedly sampling identical prompts makes prompt-cache hits an
expected artifact of our design, and per-response usage accounting
confirms it: 13.4\% of census requests reported a prompt-cache hit,
covering 10.1\% of all input tokens.
Prefix caching, however, is not a confound: it reuses the deterministic
key--value computation of an already-seen input prefix, changing the
cost and latency of a request but not the conditional distribution from
which the completion is decoded, so repetitions remain independent
draws from the same distribution. The qualitatively different
mechanism -- \emph{response-level} caching, replaying a stored
completion instead of decoding afresh -- would artificially collapse the
observed $T{=}1.0$ distribution. Within our threat model this is not a
measurement nuisance but itself an endpoint-integrity violation, and it
leaves a joint signature that the collected data expose directly:
response-variance collapse at $T{=}1.0$ co-occurring with anomalously
low and low-variance latency, which we record per response. Cells
flagged by this screen are treated as anomalies (Sec.~\ref{sec:ethics}),
not as fingerprint evidence. Running the screen over the census found
no response-caching signature: of 2{,}040 single-answer cells at
$T{=}1.0$ (peaked distributions are common and expected), only 14 also
had a median latency below half their model's median, and all of those
sat at ordinary sub-second latencies attributable to provider load
variance -- none approached the near-instant profile a served cache
would produce.

\subsection{Limitations}\label{sec:limitations}
Fingerprints drift with model updates, so reference enrollment must be
refreshed; we quantify short-horizon stability but not multi-month
drift. Verification
presumes a trusted reference deployment. Very low-traffic providers
yield few samples per provider for the robustness analysis, and a small
number of catalog-listed endpoints were not serveable during the
collection window: five returned only errors (a decommissioned
serverless deployment; two providers persistently down; two
``unknown model'' entries), and one further model's sole provider
failed systematically on every numeric prompt — all reported as
attrition rather than analyzed. Closed
vendors' family labels rely on public documentation; label noise would
depress, not inflate, our lineage results. Endpoints with mandatory
hidden reasoning are excluded: the answer emitted after a reasoning
phase samples a different generative process than the direct one-token
completion our fingerprint is defined on, and pooling the two regimes
would risk clustering by protocol rather than lineage. The criterion
is ultimately \emph{behavioral}, not declarative: collection revealed
endpoints that advertise reasoning as optional yet ignore the disable
flag (burning the token budget on a visible trace, in some cases only
at certain upstream providers), and flagship chat endpoints whose
completions consume several times more tokens than the visible
one-word answer while exposing no reasoning trace at all. For the
latter we cannot certify that the answer is a direct single-pass
sample rather than the output of a hidden deliberation phase, so we
excluded them on the same grounds rather than risk contaminating
fingerprints with post-reasoning samples; the observed hidden token
overhead is itself an ecosystem observation
(Sec.~\ref{sec:results}). The same screen operates at the response
level: a small share of completions arrived with a reasoning trace
despite the disable flag (concentrated at specific upstream
providers); such responses are coded as a separate answer class and
never enter fingerprints, and the affected model--provider
combinations were additionally screened across the entire collection:
0.76\% of census responses (2{,}486 of 326{,}047, from 14
model--provider combinations) were excluded this way. Fingerprinting the post-reasoning answer channel under a
relaxed token budget is a natural, methodologically distinct
extension. Finally, our census
covers one aggregator; the protocol is aggregator-agnostic, but
ecosystem findings may not generalize.

\section{Ethics and Responsible Disclosure}\label{sec:ethics}
Probing commercial endpoints with ordinary chat requests within paid
quotas raises no consent issues, and no personal data are involved.
Ecosystem anomalies are reported as statistical deviations between a
served distribution and a reference deployment -- not as accusations of
fraud, since benign causes (updated weights, sanctioned quantization,
caching layers) can produce deviations. The findings are not security
vulnerabilities: publication enables no attack and there is no exploit
to patch, so coordinated-disclosure norms do not apply and we did not
operate a pre-publication notification process. Accountability is
instead served by transparency: every finding is presented together
with its benign explanations, and the released raw data -- per-request
timestamps, serving provider, and generation identifiers -- allow any
named party, and any third party, to independently verify or contest
the measurements. Substantiated clarifications from affected parties
will gladly be reflected in revised versions of this article. The
released artifact enables defenders and clients to audit endpoints;
the same capability offers little offensive value, as it reveals
nothing about model internals beyond public behavior. Dual-use aspects
of shared-prior measurements (e.g., cross-model coordination) are
deferred to, and discussed in, the companion study.

\section{Conclusion}
Model identity leaks through the cheapest possible channel: the single
token a model utters when asked a trivial question. We showed that
distributions of such tokens fingerprint LLMs, reconstruct their
genealogy, and support a practical, paraphrase-robust verification
protocol whose reliability--cost curve we quantify. As inference
markets deepen and the gap between advertised and served models becomes
a routine integrity question, single-token auditing offers a
deployable, near-zero-cost forensic instrument. Future work includes
longitudinal drift tracking, watermark-aware variants, and the
companion analysis of what shared fingerprint structure implies for
tacit inter-model coordination.

\section*{Acknowledgments}
Claude Code (Anthropic) was used to provide ongoing assistance with the
study design, data collection, processing of results, and drafting of
the manuscript. All outputs were verified and corrected by the author.
The OpenRouter API credit was funded by the author's personal funds.

\bibliographystyle{IEEEtran}
\bibliography{references}

\begin{thebibliography}{10}
\providecommand{\url}[1]{#1}
\csname url@samestyle\endcsname
\providecommand{\newblock}{\relax}
\providecommand{\bibinfo}[2]{#2}
\providecommand{\BIBentrySTDinterwordspacing}{\spaceskip=0pt\relax}
\providecommand{\BIBentryALTinterwordstretchfactor}{4}
\providecommand{\BIBentryALTinterwordspacing}{\spaceskip=\fontdimen2\font plus
\BIBentryALTinterwordstretchfactor\fontdimen3\font minus
  \fontdimen4\font\relax}
\providecommand{\BIBforeignlanguage}[2]{{%
\expandafter\ifx\csname l@#1\endcsname\relax
\typeout{** WARNING: IEEEtran.bst: No hyphenation pattern has been}%
\typeout{** loaded for the language `#1'. Using the pattern for}%
\typeout{** the default language instead.}%
\else
\language=\csname l@#1\endcsname
\fi
#2}}
\providecommand{\BIBdecl}{\relax}
\BIBdecl

\bibitem{gao2025model}
\BIBentryALTinterwordspacing
I.~Gao, P.~Liang, and C.~Guestrin, ``Model equality testing: Which model is
  this {API} serving?'' in \emph{The Thirteenth International Conference on
  Learning Representations (ICLR)}, 2025. [Online]. Available:
  \url{https://openreview.net/forum?id=QCDdI7X3f9}
\BIBentrySTDinterwordspacing

\bibitem{cai2025getting}
\BIBentryALTinterwordspacing
W.~Cai, T.~Shi, X.~Zhao, and D.~Song, ``Are you getting what you pay for?
  {Auditing} model substitution in {LLM} {APIs},'' arXiv preprint
  arXiv:2504.04715, 2025. [Online]. Available:
  \url{https://arxiv.org/abs/2504.04715}
\BIBentrySTDinterwordspacing

\bibitem{zhu2025auditing}
\BIBentryALTinterwordspacing
X.~Zhu, Y.~Ye, T.~Qiu, H.~Zhu, S.~Tan, A.~Mannan, J.~Michala, R.~A. Popa, and
  W.~Neiswanger, ``Auditing black-box {LLM} {APIs} with a rank-based uniformity
  test,'' arXiv preprint arXiv:2506.06975, 2025. [Online]. Available:
  \url{https://arxiv.org/abs/2506.06975}
\BIBentrySTDinterwordspacing

\bibitem{kirchenbauer2023watermark}
\BIBentryALTinterwordspacing
J.~Kirchenbauer, J.~Geiping, Y.~Wen, J.~Katz, I.~Miers, and T.~Goldstein, ``A
  watermark for large language models,'' in \emph{Proceedings of the 40th
  International Conference on Machine Learning (ICML)}, ser. Proceedings of
  Machine Learning Research, vol. 202.\hskip 1em plus 0.5em minus 0.4em\relax
  PMLR, 2023, pp. 17\,061--17\,084. [Online]. Available:
  \url{https://proceedings.mlr.press/v202/kirchenbauer23a.html}
\BIBentrySTDinterwordspacing

\bibitem{xu2024instructional}
J.~Xu, F.~Wang, M.~D. Ma, P.~W. Koh, C.~Xiao, and M.~Chen, ``Instructional
  fingerprinting of large language models,'' in \emph{Proceedings of the 2024
  Conference of the North American Chapter of the Association for Computational
  Linguistics: Human Language Technologies (Volume 1: Long Papers)}.\hskip 1em
  plus 0.5em minus 0.4em\relax Mexico City, Mexico: Association for
  Computational Linguistics, 2024, pp. 3277--3306.

\bibitem{shao2026fitprint}
S.~Shao, H.~Zhu, Y.~Li, H.~Yao, T.~Zhang, and Z.~Qin, ``{FIT-Print}: Toward
  false-claim-resistant model ownership verification via targeted
  fingerprint,'' \emph{IEEE Transactions on Information Forensics and
  Security}, vol.~21, pp. 5612--5626, 2026.

\bibitem{uchendu2020authorship}
A.~Uchendu, T.~Le, K.~Shu, and D.~Lee, ``Authorship attribution for neural text
  generation,'' in \emph{Proceedings of the 2020 Conference on Empirical
  Methods in Natural Language Processing (EMNLP)}.\hskip 1em plus 0.5em minus
  0.4em\relax Association for Computational Linguistics, 2020, pp. 8384--8395.

\bibitem{sun2025idiosyncrasies}
\BIBentryALTinterwordspacing
M.~Sun, Y.~Yin, Z.~Xu, J.~Z. Kolter, and Z.~Liu, ``Idiosyncrasies in large
  language models,'' in \emph{Proceedings of the 42nd International Conference
  on Machine Learning (ICML)}, ser. Proceedings of Machine Learning Research,
  vol. 267.\hskip 1em plus 0.5em minus 0.4em\relax PMLR, 2025, pp.
  57\,854--57\,885. [Online]. Available:
  \url{https://proceedings.mlr.press/v267/sun25z.html}
\BIBentrySTDinterwordspacing

\bibitem{pasquini2025llmmap}
\BIBentryALTinterwordspacing
D.~Pasquini, E.~M. Kornaropoulos, and G.~Ateniese, ``{LLMmap}: Fingerprinting
  for large language models,'' in \emph{Proceedings of the 34th USENIX Security
  Symposium (USENIX Security 25)}.\hskip 1em plus 0.5em minus 0.4em\relax
  Seattle, WA, USA: USENIX Association, 2025. [Online]. Available:
  \url{https://www.usenix.org/conference/usenixsecurity25/presentation/pasquini}
\BIBentrySTDinterwordspacing

\bibitem{gubri2024trap}
M.~Gubri, D.~Ulmer, H.~Lee, S.~Yun, and S.~J. Oh, ``{TRAP}: Targeted random
  adversarial prompt honeypot for black-box identification,'' in \emph{Findings
  of the Association for Computational Linguistics: ACL 2024}.\hskip 1em plus
  0.5em minus 0.4em\relax Bangkok, Thailand: Association for Computational
  Linguistics, 2024, pp. 11\,496--11\,517.

\bibitem{renda2023can}
\BIBentryALTinterwordspacing
A.~Renda, A.~Hopkins, and M.~Carbin, ``Can {LLMs} generate random numbers?
  {Evaluating} {LLM} sampling in controlled domains,'' in \emph{ICML 2023
  Workshop: Sampling and Optimization in Discrete Space (SODS)}, 2023.
  [Online]. Available: \url{https://openreview.net/forum?id=Vhh1K9LjVI}
\BIBentrySTDinterwordspacing

\bibitem{vankoevering2024random}
\BIBentryALTinterwordspacing
K.~Van~Koevering and J.~Kleinberg, ``How random is random? {Evaluating} the
  randomness and humaness of {LLMs}' coin flips,'' arXiv preprint
  arXiv:2406.00092, 2024. [Online]. Available:
  \url{https://arxiv.org/abs/2406.00092}
\BIBentrySTDinterwordspacing

\bibitem{harrison2024comparison}
\BIBentryALTinterwordspacing
R.~M. Harrison, ``A comparison of large language model and human performance on
  random number generation tasks,'' arXiv preprint arXiv:2408.09656, 2024.
  [Online]. Available: \url{https://arxiv.org/abs/2408.09656}
\BIBentrySTDinterwordspacing

\bibitem{coronadoblazquez2025deterministic}
\BIBentryALTinterwordspacing
J.~Coronado-Bl{\'a}zquez, ``Deterministic or probabilistic? {The} psychology of
  {LLMs} as random number generators,'' arXiv preprint arXiv:2502.19965, 2025.
  [Online]. Available: \url{https://arxiv.org/abs/2502.19965}
\BIBentrySTDinterwordspacing

\bibitem{wang2025bosc}
J.~Wang, B.~Tondi, and M.~Barni, ``{BOSC}: A backdoor-based framework for open
  set synthetic image attribution,'' \emph{IEEE Transactions on Information
  Forensics and Security}, vol.~20, pp. 8043--8058, 2025.

\bibitem{zheng2026adaparse}
Y.~Zheng, Z.~Li, B.~Yu, J.~Zhou, and J.~Lu, ``{AdaParse}: Personalized
  fingerprinting for visual generative model reverse engineering,'' \emph{IEEE
  Transactions on Information Forensics and Security}, vol.~21, pp. 2682--2697,
  2026.

\bibitem{zhang2025softcontrastive}
Q.~Zhang, X.~Zhang, M.~Sun, and J.~Yang, ``A soft-contrastive pseudo learning
  approach toward open-world forged speech attribution,'' \emph{IEEE
  Transactions on Information Forensics and Security}, vol.~20, pp. 1135--1148,
  2025.

\bibitem{wei2024bdmmt}
J.~Wei, M.~Fan, W.~Jiao, W.~Jin, and T.~Liu, ``{BDMMT}: Backdoor sample
  detection for language models through model mutation testing,'' \emph{IEEE
  Transactions on Information Forensics and Security}, vol.~19, pp. 4285--4300,
  2024.

\bibitem{akata2025playing}
E.~Akata, L.~Schulz, J.~Coda-Forno, S.~J. Oh, M.~Bethge, and E.~Schulz,
  ``Playing repeated games with large language models,'' \emph{Nature Human
  Behaviour}, vol.~9, pp. 1380--1390, 2025.

\bibitem{bommasani2022picking}
\BIBentryALTinterwordspacing
R.~Bommasani, K.~A. Creel, A.~Kumar, D.~Jurafsky, and P.~Liang, ``Picking on
  the same person: Does algorithmic monoculture lead to outcome
  homogenization?'' in \emph{Advances in Neural Information Processing Systems
  35 (NeurIPS 2022)}, 2022. [Online]. Available:
  \url{https://proceedings.neurips.cc/paper_files/paper/2022/hash/17a234c91f746d9625a75cf8a8731ee2-Abstract-Conference.html}
\BIBentrySTDinterwordspacing

\bibitem{kleinberg2021algorithmic}
J.~Kleinberg and M.~Raghavan, ``Algorithmic monoculture and social welfare,''
  \emph{Proceedings of the National Academy of Sciences}, vol. 118, no.~22, p.
  e2018340118, 2021.

\bibitem{schelling1960strategy}
T.~C. Schelling, \emph{The Strategy of Conflict}.\hskip 1em plus 0.5em minus
  0.4em\relax Cambridge, MA: Harvard University Press, 1960.

\bibitem{mehta1994nature}
\BIBentryALTinterwordspacing
J.~Mehta, C.~Starmer, and R.~Sugden, ``The nature of salience: An experimental
  investigation of pure coordination games,'' \emph{American Economic Review},
  vol.~84, no.~3, pp. 658--673, 1994. [Online]. Available:
  \url{https://www.jstor.org/stable/2118074}
\BIBentrySTDinterwordspacing

\bibitem{bruckner_2026_21278557}
\BIBentryALTinterwordspacing
T.~Bruckner, ``Single-token output distributions as behavioral fingerprints of
  large language models,'' Zenodo dataset, Jul. 2026. [Online]. Available:
  \url{https://doi.org/10.5281/zenodo.21278557}
\BIBentrySTDinterwordspacing

\bibitem{bruckner_2026_21278793}
\BIBentryALTinterwordspacing
T.~Bruckner, ``Single-token output distributions as behavioral fingerprints of
  large language models --- software,'' Zenodo software archive, Jul. 2026.
  [Online]. Available: \url{https://doi.org/10.5281/zenodo.21278793}
\BIBentrySTDinterwordspacing

\bibitem{bruckner2026prereg}
\BIBentryALTinterwordspacing
T.~Bruckner, ``Single-token output distributions as behavioral fingerprints of
  large language models: model verification and tacit coordination.'' OSF
  pre-registration, Jul. 2026. [Online]. Available:
  \url{https://doi.org/10.17605/OSF.IO/KXAHM}
\BIBentrySTDinterwordspacing

\end{thebibliography}

\end{document}